\documentclass[5p]{elsarticle}
\usepackage{lineno}
\usepackage[breaklinks=true]{hyperref}
\usepackage{amsmath}
\usepackage{amssymb}
\hypersetup{colorlinks,linkcolor=red,citecolor=blue,urlcolor=blue}
\usepackage{natbib}
\bibpunct{(}{)}{;}{a}{}{,} 

\journal{New Astronomy}

\newcommand{\arcm}{\ensuremath{^\prime}}
\newcommand{\farcs}{\ensuremath{^{\prime\prime}}}
\newcommand{\ubv}{\ensuremath{UB\,V}}
\newcommand{\msun}{M$_\odot$}
\newcommand{\masyr}{\ensuremath{\rm mas\,yr^{-1}}}
\newcommand{\kms}{\ensuremath{\rm km\,s^{-1}}}
\newcommand{\logg}{$\log g$}
\newcommand{\vsini}{\ensuremath{{\upsilon}\sin i}}
\newcommand{\farcm}{\ensuremath{\overset{\prime}{.}}}

\begin{document}

\begin{frontmatter}

\title{A search for peculiar stars in the open cluster Hogg~16}


\author[label1]{Stefano Cariddi}
\author[label2]{Naira M. Azatyan}
\author[label3]{Petr Kurf\"urst}
\author[label4]{L\'ydia \v{S}tofanov\'a}
\author[label3,label8]{Martin Netopil\corref{mycorrespondingauthor}}
\cortext[mycorrespondingauthor]{Corresponding author. \copyright{2017}. This manuscript version is made available under the CC-BY-NC-ND 4.0 license http://creativecommons.org/licenses/by-nc-nd/4.0/}
\ead{mn.netopil@gmail.com}
\author[label3]{Ernst Paunzen}
\author[label5]{Olga I. Pintado}
\author[label6,label7]{Yael J. Aidelman}
\address[label1]{Dipartimento di Fisica e Astronomia ``Galileo Galilei'' - Universit\`{a} degli Studi di Padova, Vicolo dell'Osservatorio 3, 35122 Padova, Italy}
\address[label2]{Byurakan Astrophysical observatory, Byurakan 378433, Armenia}
\address[label3]{Department of Theoretical Physics and Astrophysics, Masaryk University, Kotl\'a\v{r}sk\'a 2, 611 37 Brno, Czech Republic}
\address[label4]{Astronomical Institute, Charles University in Prague, 180 00 Praha 8, V Hole\v{s}ovi\v{c}k\'ach 2, Czech Republic}
\address[label8]{Institut f\"ur Astrophysik, Universit\"at Wien, T\"urkenschanzstra{\ss}e 17, A-1180 Wien, Austria}
\address[label5]{Centro de Tecnolog\'ia Disruptiva, Universidad de San Pablo Tucum\'an, San Pablo, Tucum\'an, Argentina}
\address[label6]{Instituto de Astrof\'isica de La Plata, CCT La Plata, CONICET-UNLP, Paseo del Bosque S/N, B1900FWA, La Plata, Argentina}
\address[label7]{Departamento de Espectroscop\'ia, Facultad de Ciencias Astron\'omicas y Geof\'isicas, Universidad Nacional de La Plata (UNLP), Paseo del Bosque S/N, B1900FWA, La Plata, Argentina}

\begin{abstract}
The study of chemically peculiar (CP) stars in open clusters provides valuable information about their evolutionary status. Their detection can be performed using the $\Delta a$ photometric system, which maps a characteristic flux depression at $ \lambda \sim 5200$ \AA. This paper aims at studying the occurrence of CP stars in the earliest stages of  evolution of a stellar population by applying this technique to Hogg~16, a very young Galactic open cluster ($\sim 25$\,Myr).
We identified several peculiar candidates: two B-type stars with a negative $\Delta a$ index (CD\,$-$60~4701, CPD\,$-$60~4706) are likely emission-line (Be) stars, even though spectral measurements are necessary for a proper classification of the second one; a third object (CD\,$-$60~4703), identified as a Be candidate in literature, appears to be a background B-type supergiant with no significant $\Delta a$ index, which does not rule out the possibility that it is indeed peculiar as the normality line of $\Delta a$ for supergiants has not been studied in detail yet. A fourth object (CD\,$-$60~4699) appears to be a magnetic CP star of 8 M$_\odot$, but obtained spectral data seem to rule out this hypothesis.  Three more magnetic CP star candidates are found in the domain of early F-type stars. One is a probable nonmember and close to the border of significance, but the other two are probably pre-main sequence cluster objects. This is very promising, as it can lead to very strong constraints to the diffusion theory.
Finally, we derived the fundamental parameters of Hogg~16 and provide for the first time an estimate of its metal content.

\end{abstract}


\end{frontmatter}


\section{Introduction}

Chemically peculiar (CP) stars of the upper main sequence are B to early F-type stars that show a variety of elemental abundance pattern. \citet{preston74} divided them into four groups: Am
stars (CP1), Bp/Ap stars (CP2), HgMn stars (CP3), and He peculiar
stars (CP4). Two groups (CP2 and CP4) also show strong and ordered
magnetic fields up to about 30\,kG, and in the following, we refer to them as magnetic chemically peculiar (mCP) stars. The study of mCP stars (and all other star groups as well) in open clusters provides valuable information about their evolutionary status, because open clusters provide strong constraints to the stellar age and other parameters. For example, already \citet{young73} conducted a spectroscopic search for these objects in a number of clusters to investigate their incidence. Later, \citet{hartoog77} made use of the benefits of open clusters to study the rotation and magnetic braking of mCP stars. More recent works in this respect are for example by \citet{bailey14} and \citet{landstreet07} about the  evolution of elemental abundances and magnetic field strength of mCP stars, or the study by \citet{fossati08} of Am stars in the Praesepe cluster.

Thorough statistical investigations require large samples of well confirmed cluster CP stars to cover a wide mass and age range. It is therefore essential to detect additional (preferably bright) candidates for detailed follow-up observations. However, a careful evaluation of the cluster membership is inevitable and might be complicated by the absence of datasets with complete kinematic information. We note that field mCP stars are also helpful to improve our knowledge of their evolutionary status, although one can not directly constrain the stellar age. Large samples with a measured $Hipparcos$ parallax were analysed for example by \citet{gomez98} and \citet{hubrig00}, or recently by \citet{netopil17} based on a mix of $Hipparcos$ and $Gaia$ data.

The $\Delta a$ photometric system \citep{maitzen76} allows an efficient detection of mCP star candidates because of the characteristic flux depression at $\lambda$\,5200\,\AA\ \citep{kodaira69,kupka03}. The iron abundance was found as the main contributor into the flux depression for the whole range of
effective temperatures, while Cr and Si are important primarily for low effective temperatures \citep{khan07}. By measuring the flux depression and the adjacent regions, peculiar objects can be easily separated from normal stars. \citet{paunzen05c} have shown that up to 95\,\% of mCP stars can be detected by $\Delta a$. Most objects display positive $\Delta a$ values up to 60\,mmag, though extreme values of the order of 100\,mmag are also found \citep{netopil07}. The class of $\lambda$\,Bootis stars \citep[see e.g.][]{Murphy17} show moderate
to strong underabundances of Fe-peak elements and have a detection efficiency comparable to the magnetic stars; almost two third can be detected at a limit of $-$10\,mmag. Only extraordinary strong peculiar objects of other non-magnetic CP groups are outstanding in $\Delta a$. For example, the detection capability of Am stars
is only 17 percent for a limit of +10\,mmag in $\Delta a$ \citep{paunzen05c}. Although variability or binarity might interfere the results, the photometric system can be considered as an efficient pre-selection tool in particular for mCP stars. The system was not only applied to Galactic field stars, open clusters \citep{netopil07} and globular clusters \citep{paunzen14b}, but also to clusters in the Large Magellanic Cloud \citep{paunzen06b}, where photometric mCP star detections are also confirmed by spectroscopy \citep{paunzen11}. 

This work presents a $\Delta a$ study of the Galactic open cluster Hogg~16, a very young aggregate, which therefore adds an important contribution to a study of the occurrence of CP stars at the earliest stage of stellar evolution. A better knowledge of the  evolutionary status of mCP stars is of importance to understand the mechanism of the  magnetic field generation, the evolution of the magnetic field strength and its geometry, or to provide constraints for the diffusion theory.

The paper is arranged as follows: Section \ref{obsdata} describes the data and the reduction procedure, in Sect.~\ref{param} we derive the cluster parameters, Sect.~\ref{cpcandidates} discusses the detected chemically peculiar objects, and finally Sect.~\ref{summary} concludes the work.

\section{Observations and data reduction}
\label{obsdata}
The photometric observations of Hogg~16 were performed on 2004 June 15  with the EFOSC2 instrument, installed on the 3.6\,m telescope at ESO - La Silla within the program 073.C-0144(A), and the target field was centred on the main concentration of stars in the cluster area (J2000 RA 13:29:18, DEC $-$61:12:00). The field-of-view is about 5\farcm2 $\times$ 5\farcm2, and the 2\,$\times$\,2 binning mode results in a resolution of 0.31\farcs\,pixel$^{-1}$. Thus, we cover almost the complete cluster area if adopting a diameter of 6\arcm\ as listed in the updated open cluster catalogue
by \citet[][version 3.5]{dias02}.  We used a $\Delta a$ filter set with the following characteristics: $g_1$ ($\lambda_c$ = 5007\AA, FWHM = 126\AA, $T_P$ = 78\%), $g_2$ (5199, 95, 68), 
and $y$ (5466, 108, 70). To cover the broadest possible magnitude range with a good signal-to-noise ratio and without saturation, a set of 10 observations in each filter with short (5\,seconds) and longer exposures (70--100\,seconds) were obtained, resulting in 60 scientific frames in total. The basic CCD reductions (bias subtraction and flatfield correction) and point-spread-function fitting were carried out with standard IRAF V2.16 routines. The method of calculating the normality line (the reference $a$ index of apparently normal type stars), deriving the errors, and the calibration of our ($g_1 - y$), and $y$ measurements is identical to that of previous works \citep[see e.g.,][]{netopil05,netopil07,paunzen06} and
we refer to them for more details.

We used the ($b-y$)/$V$ photometry by \citet{mcswain05} for the calibration of the ($g_1 - y$)/$y$ data to obtain standardised photometry.  The transformation coefficients are given in Table~\ref{tab:regression}. For completeness, also the results for the normality line (using stars earlier than spectral type F2) and of the linear fit to the broadband Johnson ($B-V$) colours by \citet{vazquez91} are listed. Our data provide a two magnitudes fainter limit than the last optical study of the target \citep{mcswain05}, which allows us to investigate also lower mass stars in the open cluster. We also queried for data in the $I$ band from the Denis survey \citep{epchtein97} and $JHK_{\rm s}$ 2MASS photometry \citep{skrutskie06}, but we only adopted the most reliable 2MASS data that belong to the quality category~A. 

Furthermore, we queried for proper motion data in the UCAC5 \citep{ucac5}, UCAC4 \citep{ucac4}, and PPMXL \citep{ppmxl} catalogues and derived kinematic membership probabilities using a completely non-parametric approach \citep[][]{galadi98,balaguer04}. We note that a comparison of the individual results does not show any correlation of the membership probabilities. The median proper motions agree very well, thus we attribute this finding to the errors of the measurements. These are taken into account in the membership analysis, but each catalogue shows  different errors for individual objects. By considering the proper motion errors we selected kinematic 3$\sigma$ members based on the currently most accurate UCAC5 data, followed by data from UCAC4 and PPMXL, respectively, if not covered by the primary source. Photometric data complement the member star selection by identifying objects that deviate from the cluster sequence in the colour-magnitude diagrams (CMD). For 72 stars out of 150 measured objects in the field we finally define a probable cluster membership.   

Currently available radial velocity measurements are not helpful to constrain membership to this clusters. These are only available for four stars in our sample and mean cluster velocities differ significantly: \citet{kharch05} list $-36 \pm 7 $\,\kms\ based on four stars and \citet{conrad14} derived $-51 \pm 2$\,\kms\ using data of two stars. Unfortunately, some spectra which are at our disposal (see also discussion in Sect. \ref{cpcandidates}) are either of too low resolution or without sufficient standard star observations and do not offer accurate radial velocities for a few additional stars. A kinematic membership determination  based only on proper motion data introduces  limitations for distant clusters, which evidently do not show large differences in motion between the field and cluster population. This problem is nicely demonstrated for example in the work by \citet{sanner01}. Future data releases of the $Gaia$ mission will certainly provide the basis for detailed membership analyses of star clusters using precise proper motions, radial velocities, and parallaxes.

\begin{table}
\caption{Regression coefficients for the normality line and colour transformations based on a number (N) of stars. The errors of the last significant digits are given in parentheses.} 
\label{tab:regression} 
\centering 
\begin{tabular}{l l l} 
\hline
 	& regression & N\\ 
\hline 
$a_0$ & 0.013(2) + 0.195(26).($g_1 - y)$ & 21 \\
$V$ & 2.116(48) + 0.999(4).$y$ & 28 \\
($b-y$) & 0.320(6) + 1.652(32).($g_1 - y)$ & 28 \\
($B-V$) & 0.395(15) + 2.753(75).($g_1 - y)$ & 19 \\

\hline 
\end{tabular}
\end{table}

\section{Parameters of Hogg~16}
\label{param}

\begin{table}
\caption{Literature results for the open cluster Hogg~16.} 
\label{tab:literature} 
\centering 
\begin{tabular}{rrcl} 
\hline
log\,\textit{t}	& d[pc]  & $E(B-V)$ & Reference\\ 
\hline 
--  & 1810 & 0.42 & \citet{moffat73} \\
$<$8.70 & 603 & 0.36 & \citet{fenkart77} \\
7.41 & 2130 & 0.44 & \citet{vazquez91} \\
7.40 & 1905 & 0.45 & \citet{dambis99} \\
7.05 & 1585 & 0.41 & \citet{loktin01} \\
7.26 & 1585 & 0.41 & \citet{kharch05} \\
7.05 & 1585 & 0.41 & \citet{mcswain05} \\
7.95 & 2059 & 0.41 & \citet{kharch13} \\
7.04 & 3172 & 0.63 & \citet{aidelm15} \\
$\leq$7.40 & 2040 & 0.45 & this study \\
\hline 
\end{tabular}
\end{table}

Several studies of Hogg~16 are available in the literature, and we list an overview of the open cluster parameters in Table~\ref{tab:literature}. We note that the reddening values by \citet{fenkart77} and \citet{mcswain05} are based on the $RGU$ and Str\"omgren system, respectively. These were transformed to $E(B-V)$ using the colour excess ratios by \citet{steinlin68} and \citet{mccall04}. Some studies list  identical parameter values, thus these were probably adopted from the work by \citet{loktin01}; see also the discussion by \citet{npc15}. Two studies \citep{fenkart77,kharch13} list a much older age than the others. The first reference was certainly limited by the data and evolutionary models available at that time. The result by the latter reference is based on 2MASS data and is part of a large compilation of open cluster parameters. They analysed about 3000 open clusters, thus only a little time can be probably spent for a detailed check of each object and discrepant parameter results are often reported in the literature \citep[see e.g. discussion by][]{npc15}.  Nevertheless, most works quote a cluster age younger than about 30\,Myr, and a distance close to 2\,kpc.  \citet{aidelm15} derived the largest distance and the highest reddening of Hogg~16 by using the Barbier-Chalonge-Divan (BCD) spectrophotometric system and low-resolution spectroscopy obtained with the 2.15\,m telescope at the Complejo Astron\'omico El Leoncito (CASLEO), San Juan, Argentina. They identified two populations in the direction of the cluster: a nearby young group ($<$\,40\,Myr) at a distance of 605\,pc \citep[in agreement with the result by][]{fenkart77} and a more distant group with the parameters listed in Table~\ref{tab:literature}. The latter population is most likely the cluster which was identified by the other references in Table~\ref{tab:literature}.

Another open cluster (Collinder~272) is located about 11\arcm\ away from Hogg~16. The cluster parameters are close to the ones of Hogg~16, thus \citet{vazquez97} noted that they could represent an example of a sequential formation. This cluster was already a target of the $\Delta a$ study by \citet{paunzen02}, but they have not detected any chemically peculiar object.

We applied the method by \citet{poehnl10} to estimate the parameters of Hogg~16. It is based on photometric data and uses isochrones that are normalised to the zero-age main-sequence (ZAMS). In the following, we refer to it as the DG (differential grid) method. In an iterative way the cluster parameters distance, age, reddening, and metallicity can be derived. A comparison with metallicities derived from high-resolution spectroscopic data results in excellent agreement with a scatter of only about 0.05\,dex \citep{heiter14,netopil15}. The method was already applied to almost all open clusters of the $\Delta a$ survey, confirming a homogeneous scale of the cluster parameters \citep{netopil13}. 

\begin{figure}
\centering
\resizebox{\hsize}{!}{\includegraphics{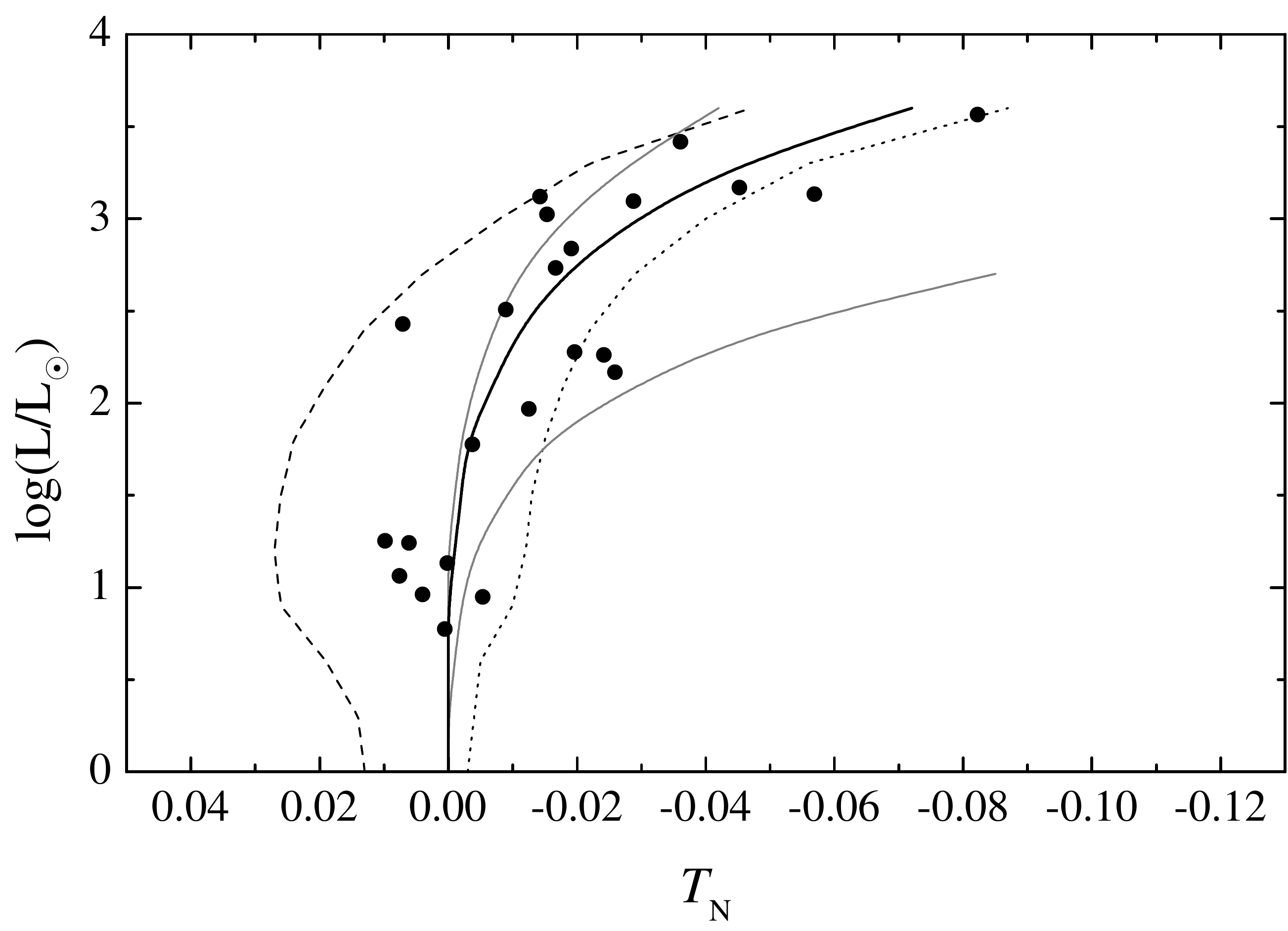}}
\caption{Hogg~16 analysed with the method by \citet{poehnl10}. The solid black line represents the solar metallicity isochrone for log\,\textit{t} = 7.4. The dashed and dotted lines represent the underabundant (Z=0.01) and overabundant (Z=0.03) isochrone of the same age, respectively. As comparison, we include solar metallicity isochrones with an age of log\,\textit{t} = 7.2 and 8.0 (grey lines). $T_{\rm N}$ is the temperature difference in dex between the star and the ZAMS at solar metallicity.}
\label{fig:dgplot}
\end{figure}

For Hogg~16 we use the \ubv\ data by \citet{vazquez97} to estimate the mean reddening of the B-type stars with the Q-method \citep{johnson58}, resulting in $E(B-V) = 0.45 \pm 0.05$\,mag based on 18 objects. The open cluster shows a slight differential reddening, a typical characteristic in young areas. Thus, in particular for the B-type cluster stars, the effective temperature as an input to the DG method will be uncertain if adopting a mean reddening. We therefore used the temperature calibration for the reddening-free Q-index by \citet{netopil13}. Additionally, we selected some fainter (cooler) stars with $(b-y)$, $(V-K_{\rm s})$, or $(V-I)$ photometry. These data were transformed to effective temperatures by applying the mean cluster reddening given above and the respective reddening ratios and temperature calibrations compiled by \citet{netopil13}. Finally, mean temperatures were derived for the cluster stars. However, we excluded objects cooler than about 7000\,K (the cut-off corresponds to early F-type stars). At the young age of Hogg~16 these objects are probably still on the pre main-sequence (PMS) and their inclusion would influence the DG method which relies only on the main-sequence. Figure\,\ref{fig:dgplot} shows the solution for the cluster based on 23 normal type main-sequence member stars. The final cluster parameters of Hogg~16 apart from the reddening that we defined above are: log\,\textit{t} = 7.4, $(m-M)_0$ = 11.55\,mag, and Z = 0.021 $\pm$ 0.007. The derived Z value represents an iron abundance of [Fe/H] = 0.01 $\pm$ 0.16\,dex using the transformation suggested by \citet{netopil15}. We note that the given age should be considered as an upper limit based on few higher mass main-sequence objects. With our limited data it is usually difficult to pin down the age of such young aggregates, because there are no highly evolved stars visible. Furthermore, an age spread owing to sequential star formation, for example, might be present as well.

\begin{figure*}
\centering
\resizebox{\hsize}{!}{\includegraphics{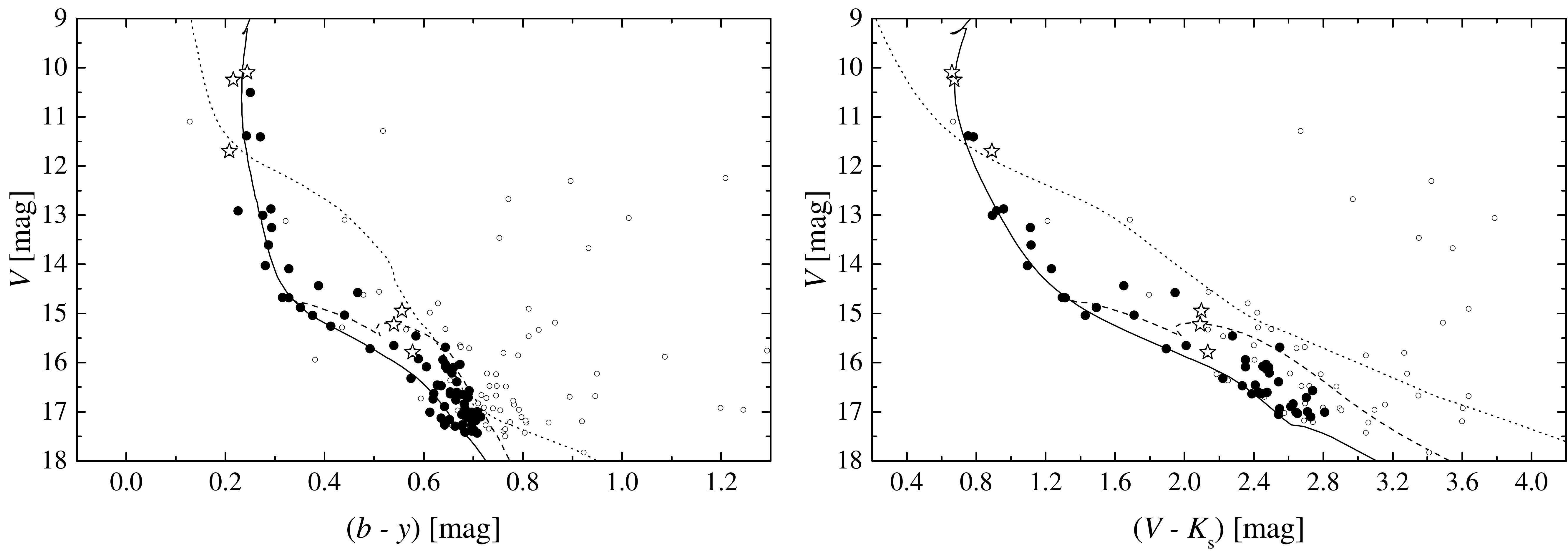}}
\caption{Colour-magnitude diagrams of stars in the area of Hogg~16. Filled black circles represent the identified member stars, smaller open circles are probable non-members, and the star symbols the identified peculiar objects. The isochrone by \citet{bressan12} for solar metallicity and the derived cluster parameters is overplotted as a solid line. The PMS isochrone part for 10\,Myr is shown as dashed line, and we indicate the closer population that was identified by \citet{aidelm15} with the dotted line isochrone. }
\label{fig:cmd}
\end{figure*}

We also show the results for the cluster in the conventional way, thus by fitting isochrones \citep{bressan12} for solar metallicity to the CMDs (see Fig.~\ref{fig:cmd}). By adopting the distance modulus and age that we derived using the DG method, we verify the reddening also in other colours and estimate $E(b-y) = 0.32$\,mag and $E(V-K_{\rm s}) = 1.34$\,mag. The use of the extinction ratios by \citet{mccall04} results in $E(B-V)=0.43$\,mag and $E(B-V)=0.49$\,mag for the $(b-y)$ and $(V-K_{\rm s})$ colours, respectively. Furthermore, we estimated $E(V-I) = 0.60$\,mag, which corresponds to $E(B-V)=0.44$\,mag. The colour excesses suggest a normal reddening law ($R_V \sim 3.1$) for the cluster, in agreement with the result by  \citet{vazquez91}.  We note that the cluster membership is more uncertain for the fainter stars. Without additional information (for example radial velocities) it is difficult to distinguish between field stars, cluster PMS objects, member stars in binaries, or somewhat higher reddened cluster stars. Sequential star formation might contribute to some additional broadening of the cluster sequence in the CMD as well. We therefore also show the PMS isochrone for 10\,Myr in Figure~\ref{fig:cmd}. We note that the isochrones by \citet{bressan12} already include the PMS stage. The two isochrones cover the distribution of the stars in the CMD. However, based on the available data we are not able to conclude about a sequential star formation in the cluster, because a large fraction of binary stars could result in a similar appearance by a maximum shift of 0.75\,mag to brighter magnitudes. As already mentioned, \citet{aidelm15} identified another young but much closer group of stars, and we show the isochrone for this population with the parameters derived by this reference in Fig.~\ref{fig:cmd}. Our covered field is probably small enough that the closer and more extended group does not influence our results. There is only one star in common with \citet{aidelm15} that belongs to the close group (CPD\,$-$60\,4698, the bluest object in the left panel of Fig.~\ref{fig:cmd}). Furthermore, the position of the isochrones indicates that there is indeed very little interference with the distant Hogg~16 population. 

\subsection{The relationship of Hogg~16 with Collinder~272}
\label{sect:cluster_relation}
The parameters of Hogg~16 agree very well with the somewhat more populous cluster Collinder~272, which was already analysed with the DG method: log\,\textit{t} = 7.3, $(m-M)_0$ = 11.75\,mag, $E(B-V)= 0.43$\,mag, and [Fe/H] = 0.03\,dex \citep{netopil13,netopil15}. The distances differ by only 200\,pc ($\sim$\,10\,\%), which is well within typical error ranges even among homogeneous approaches \citep{npc15}. If we place both clusters at the same distance ($\sim$\,2\,kpc), the angular separation of the cluster centres (11\arcm) corresponds to only about 6\,pc. Thus, a physical or evolutionary relationship between the two clusters is most likely as proposed by \citet{vazquez97}. These authors derived a mean age of 13\,Myr for Collinder~272, which is close to the result listed above, and they note that the bright stars of Hogg~16 have probably triggered the star formation in Collinder~272.

To further evaluate the results for the two clusters we make use of the recent results of the $Gaia$ satellite mission \citep{gaiadr1}. We queried for DR1-TGAS data in not overlapping regions around the two clusters and identified five and nine stars in Hogg~16 and Collinder~272, respectively, that are most likely members according to their colours, proper motion, and parallax. We derived a mean parallax of $0.49 \pm 0.20$\,mas for Hogg~16 and $0.48 \pm 0.14$\,mas for Collinder~272, both in very good agreement with the photometric results. However, we note that an average does not lead to a gain in precision, there is still a systematic uncertainty of 0.3\,mas \citep{gaiadr1}. Thus, the distances based on $Gaia$ DR1 range from about 1.3\,kpc to more than 5\,kpc. The proper motion of the two clusters agrees within the errors as well. We obtain a mean motion $\mu_{\alpha*}/\mu_{\delta}$ of $-3.90 \pm 0.61$/$-1.85 \pm 0.92$\,\masyr\ and $-3.52 \pm 0.82$/$-2.40 \pm 0.51$\,\masyr\ for Hogg~16 and Collinder~272, respectively.

We also investigate a 60$'$ wide region around the two clusters and notice an indication of a bimodal parallax distribution, one peak mostly caused by the two clusters and another one at $1.38 \pm 0.12$\,mas. The latter might be related to the close group mentioned by \citet{aidelm15}. However, we are not able to identify a common proper motion of the closer stars.

\section{Chemically peculiar stars in Hogg~16}
\label{cpcandidates}

\begin{figure}
\centering
\resizebox{\hsize}{!}{\includegraphics{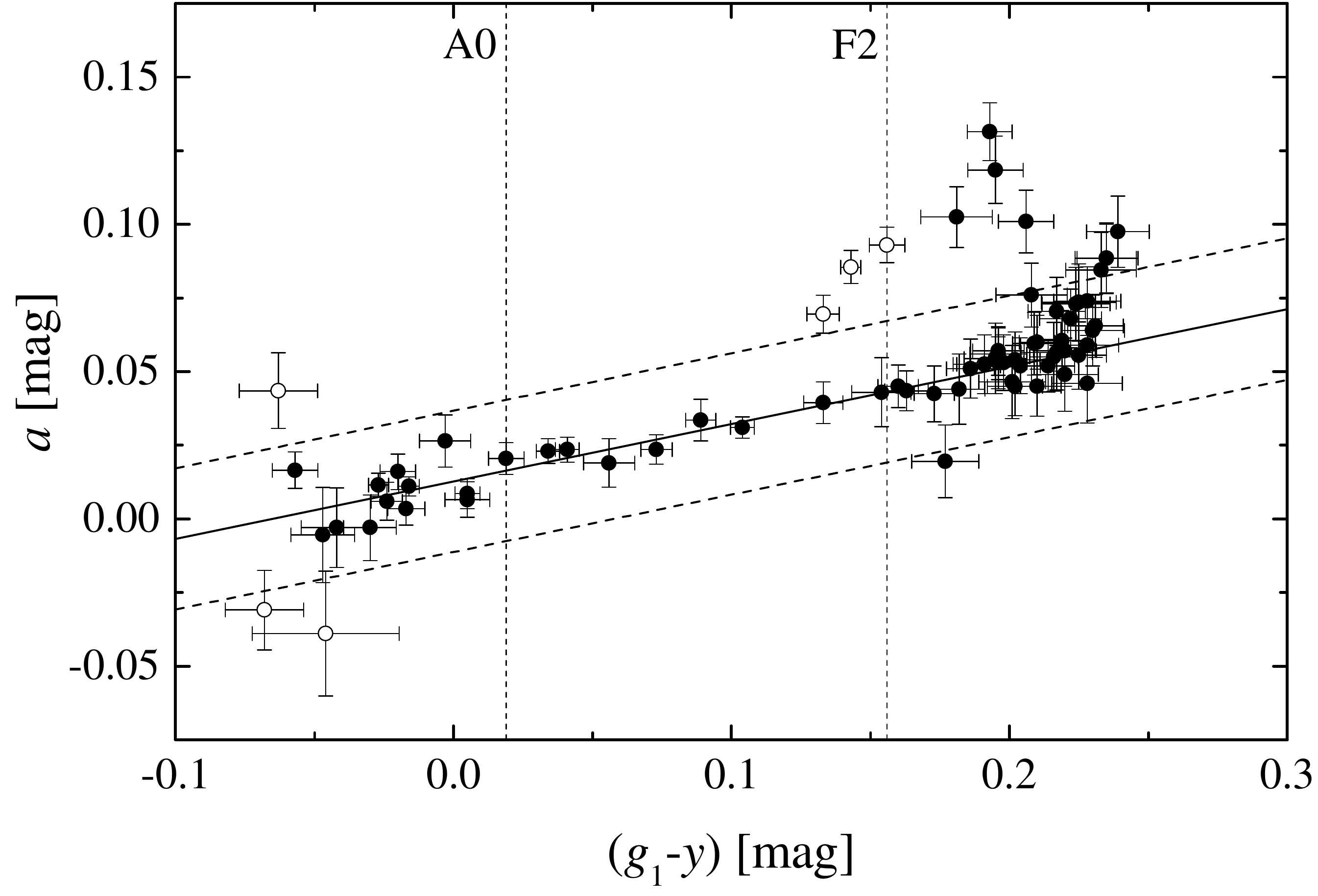}}
\caption{Diagnostic $\Delta a$ diagram for the member stars of the open cluster Hogg~16. The normality line and the three $\sigma$ confidence interval are shown as solid and dashed lines, respectively. The vertical dashed lines indicate the colour of the spectral types A0 and F2. Open circles show the chemically peculiar candidates that we discuss in the text.}
\label{fig:deltaa}
\end{figure}

We used our photometric data to identify chemically peculiar candidate stars in the cluster Hogg~16. Figure  \ref{fig:deltaa} shows the diagnostic $\Delta a$ diagram of the apparent member stars, the normality line, and the $3\,\sigma$ confidence interval. For better guidance, we also included lines that correspond to spectral types A0 and F2, respectively. The latter is the cool limit where classical chemically peculiar stars are in general still found. Table~\ref{tab:parameters} lists the parameters of the stars along with the kinematic membership flag based on proper motion data. All photometric data of the open cluster are available in electronic form at the CDS; an excerpt is shown in Table \ref{tab:allstars}.

\subsection{Emission-line stars}
Two B-type stars are located below the normality line, thus showing negative $\Delta a$ indices. This is an indication of emission and the stars can be in general classified as Be stars. For the object CD\,$-$60\,4701 the following spectral classifications are available in the literature: B1.5\,V:ne \citep{garrison77}, and B3\,V \citep{fitzg79}. \citet{aidelm15} list a spectral type of B1\,III and note that it is a Balmer emission-line object owing to the presence of a second Balmer discontinuity. The star was also classified as possible Be star by \citet{mcswain05} based on H$\alpha$ photometry. For the other Be star candidate (CPD\,$-$60\,4706) there are neither spectral types available in the literature, nor the star shows signs of H$\alpha$ emission \citep{mcswain05}. More observations, in particular spectroscopic data are needed, because Be stars can show strong variations in the spectrum, and change between their Be and shell phase \citep[see e.g.][]{mcswain09}. Available \ubv\ photometry for these two objects indicate a reddening that is in very good agreement with the cluster mean \citep[see][]{vazquez91}. Finally, the $Gaia$ data (parallax and proper motion) allow us to conclude a definite cluster membership. We note that another listed Be star is in our field of view \citep[CPD\,$-$60\,4703;][]{fitzg79}. \citet{aidelm15} classify it as emission-line star of spectral type B0\,Ib and it is therefore not a classical Be star. Although the $Gaia$ data do not rule out a membership, the object is most likely a non-member of the cluster. Both, the high reddening \citep[$\sim$\,1\,mag;][]{fitzg79,aidelm15} and the absolute magnitude of supergiants \citep{wegner06} indicate a background position. If we take the higher reddening into account, the star does not show up in $\Delta a$. Though, the characteristic of the normality line for supergiants is not studied yet in detail.

\begin{table*}
\caption{Parameters of the candidate chemically peculiar stars. We also list the kinematic membership flag in $\sigma$ based on the UCAC5/UCAC4/PPMXL data.} 
\label{tab:parameters} 
\centering 
\begin{tabular}{l c c c c c c c} 
\hline
No. (Webda \#) / ID	& RA  & DEC  & V  & $\Delta a$  & ($b-y$)$_0$  & $M_V$ & membership \\ 
	& [deg] & [deg] & [mag] & [mmag] & [mag] & [mag] & $\sigma$\\ 

\hline 
37 (--) & 202.3946417 & $-$61.2064550 & 15.22 & +31 & +0.21 & +2.28 & 3/2/1\\
43 (60) / CPD\,$-$60 4706 & 202.3777375 & $-$61.2032361 & 11.70 & $-$30 & $-$0.09 & $-$1.11 & 2/1/2\\ 
46 (21) / CD\,$-$60 4699 & 202.2955583 & $-$61.2003700 & 10.25 & +43 & $-$0.09 & $-$2.59 & --/1/1\\
53 (14) / CD\,$-$60 4701& 202.3928875 & $-$61.1933342 & 10.10 & $-$43 & $-$0.07 & $-$2.77 & 2/1/1 \\
84 (258) & 202.3280792 & $-$61.1790592 & 14.95 & +45 & +0.22 & +2.00 & 2/1/3 \\
97 (--) & 202.3052083 & $-$61.1670183 & 15.79 & +50 & +0.24 & +2.84 & 2/0/0 \\

\hline 
\end{tabular}
\end{table*}

\subsection{The photometric mCP star candidate CD\,$-$60\,4699}

One B-type star (CD\,$-$60\,4699) is located well above the normality line in Fig.~\ref{fig:deltaa}. The position in the $\Delta a$ diagram and in the CMD indicates that it is probably a mCP star candidate of about 8\,\msun. Thus, it might belong to the most massive mCP stars, the helium peculiar objects. \citet{fitzg79} list a spectral type of B2\,III, but the position of the star in the CMD does not support this luminosity class. We note that in the literature mCP stars are often misclassified as giants owing to the slow rotation and the resulting sharp spectral lines. Recently, \citet{aidelm15} obtained a spectral type of B2VI: based on the applied BCD system. All listed astrophysical  parameter values for this star are flagged as uncertain by the authors, which could be caused by the chemically peculiar characteristic. \citet{zidale07} show for example that the BCD system fails in case of He-strong stars. Larger He/H ratios produce smaller Balmer discontinuities, which results in an overestimation of temperature.

\begin{figure}
\centering
\resizebox{\hsize}{!}{\includegraphics{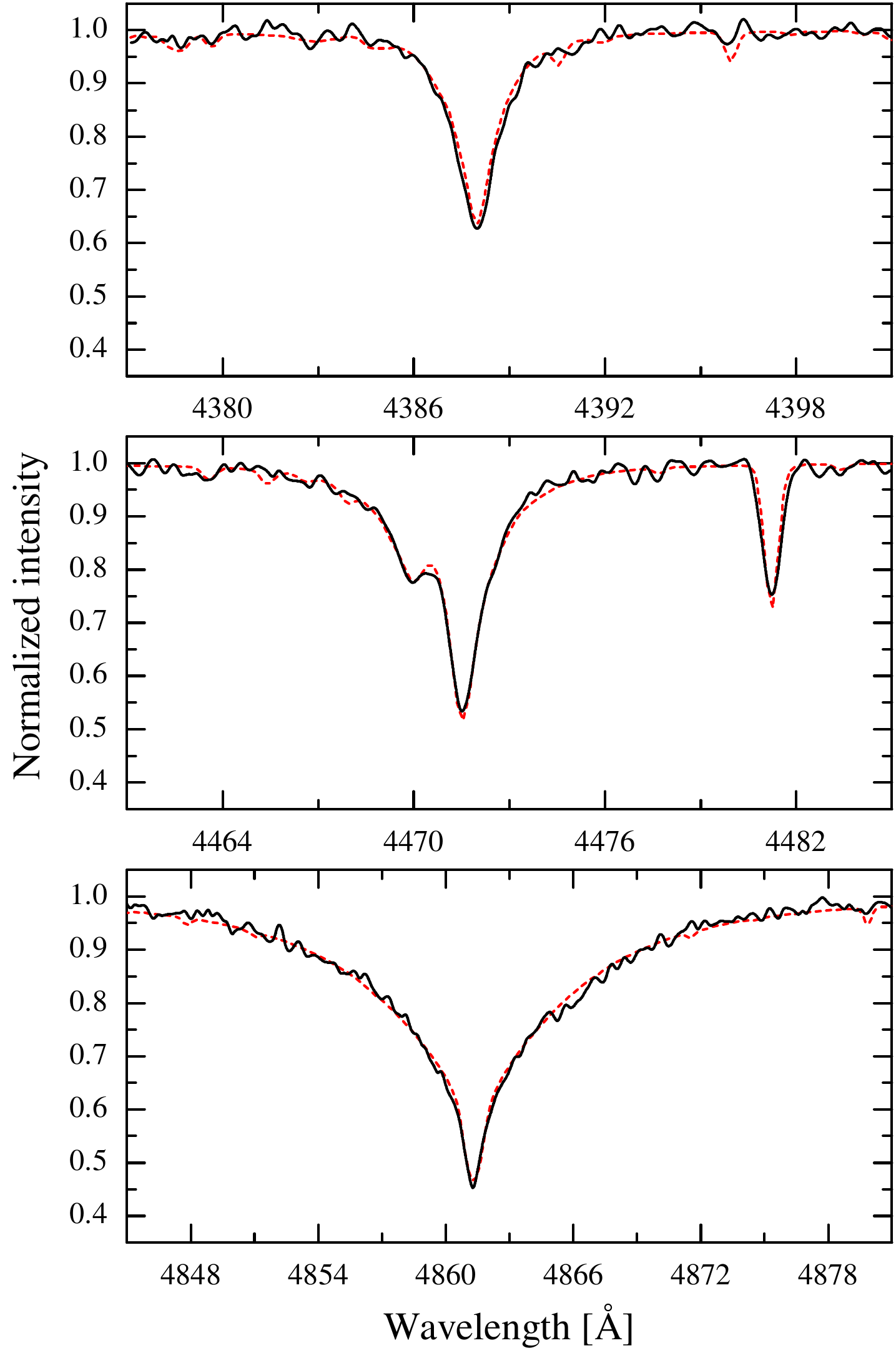}}
\caption{Details of the medium resolution spectrum of CD\,$-$60\,4699, showing the spectral lines He\,{\sc i} $\lambda$4387\,\AA, He\,{\sc i} $\lambda$4471\,\AA\ / Mg\,{\sc ii} $\lambda$4481\,\AA, and H$\beta$. A synthetic spectrum for $T_{\rm eff}=22000$\,K, \logg=4.0, and [$M/H$]=0.0 is shown as dashed line.}
\label{fig:spectrum}
\end{figure}

We therefore investigate the spectrum used by \citet{aidelm15}. However, beside a noticeable strong C\,{\sc ii} line at $\lambda$4267\,\AA\ the data appear inconclusive, probably owing to the spectral resolution (4.53\,\AA\ every two pixel). We obtained an additional spectrum at the CASLEO observatory on 2016 January 27 with the same configuration, but unbinned and slightly shifted to the red. Furthermore, we observed the star at this observatory also with higher resolution (R $\sim$ 12\,500) on 2016 March 30. Figure \ref{fig:spectrum} shows portions of the medium resolution spectrum along with the best fitting synthetic spectrum. The latter was computed using the program \texttt{SPECTRUM} \citep{gray94} and an ATLAS9 model atmosphere \citep{castelli04} for $T_{\rm eff}=22000$\,K, \logg = 4.0, [$M/H$] = 0.0, and a microturbulent velocity of 2\,\kms. We notice very sharp spectral lines, indicating a \vsini\ $\lesssim 20$\,\kms. Slow rotation is a well known characteristic of mCP stars, but the star appears normal also if we compare the equivalent widths of helium lines with results for B-type main-sequence stars \citep[][]{leone98}.

Unfortunately, the medium resolution spectrum starts at a somewhat redder wavelength than the previously noticed strong C\,{\sc ii} line, but in the new low resolution data the line appears normal. There is certainly still the possibility that the object was observed at mistimed phases, but based on the available data we have to classify it as a normal B2V star. We also note that it was identified as double star based on \textsc{Hipparcos}/$Tycho$-2 data \citep{fabricius02,leeuwen07}, but it is unfortunately not included in the first $Gaia$ data release. The positive $\Delta a$ detection might be also influenced by variability. For example, $\beta$ Cephei pulsation is possible in early B-type stars. Most objects of this group are also slow rotators \citep{stankov05}, thus the low \vsini\ value of the star does not contradict such a classification.  The object is covered by `The All Sky Automated Survey' \citep[ASAS,][]{pojmanski97}, which provides continuous photometric monitoring of the whole sky. However, the low spatial resolution of the survey results in blends with another nearby bright star. Thus, the data are not well suited to study the photometric variability. Beside new photometric measurements also some additional spectroscopic or even polarimetric data are needed for a final conclusion about the nature of this object and to investigate the reasons for the spectroscopic non-confirmation as mCP star. These will help either to explain the $\Delta a$ detection as a result of binarity or variability, but also to verify if the current spectroscopic observations were obtained at a mistimed phase.

\subsection{Cooler type mCP candidates}

Three mCP candidates are found in the domain of early F-type stars, though one object (No.\,37) is very close to the border of significance. Another object (No.\,97) lies at the F2 spectral type borderline in Fig~\ref{fig:deltaa}, but only the recent and most accurate UCAC5 proper motion data indicate a probable kinematic membership. Two objects (No.\,37 and 84) deviate significantly from the cluster sequence and are located close to the 10\,Myr PMS isochrone. These, and some other `normal' type stars, are therefore somewhat younger than the age inferred from the higher mass stars. In Section \ref{sect:cluster_relation} we discussed the possible relationship of Hogg~16 with Collinder~272. Thus, we either see a scenario of mutually triggered star formation, or these stars are members in the outskirts of the younger cluster Collinder~272. Note that both open clusters show comparable proper motions. Upcoming $Gaia$ data releases will probably allow us a more detailed analysis of the cluster areas and their member stars.

We investigated the spectral energy distribution (SED) of all three mCP star candidates using the fitting tool by \citet{robi07} and the VO Sed Analyzer \citep[VOSA;][]{bayo08}. Both tools allow to set interstellar extinction as the free parameter. For the stars No.\,84 and 97 the results agree well with the cluster reddening, but for the star No.\,37 the estimated interstellar extinction ranges from about 0.6\,mag to 0.9\,mag (the higher value obtained by VOSA). These results make the detection of the star No.\,37 as mCP object uncertain because of two reasons: first, the low reddening places the object more distant from the cluster sequence, which would indicate a non-membership. Second, the reddening correction in $\Delta a$ \citep{paunzen05,paunzen14} results in an insignificant value of the peculiarity index. The UCAC5 data for the star also show the lowest kinematic membership probability among the covered peculiar candidates.  However, additional photometric data in the $U$ band at least are needed to obtain a better coverage of the SED. We note that data from the Wide-field Infrared Survey Explorer \citep[WISE;][]{cutri13} show an IR-excess for the stars No.~84 and 97 that starts at the W3 (11.6\,$\mu \rm m$) band. For the remaining star, the WISE data are uncertain in the two reddest filters owing to S/N ratios well below 2. The presence of an IR-excess is an indication for circumstellar material around a young (PMS) object. The detection of definite PMS mCP stars will provide strong constraints for the diffusion theory. The stars No.\,84 and 97 are certainly the most promising objects for follow-up observations.  \citet{netopil14} presented a spectroscopic study of a PMS mCP candidate in the young cluster Stock~16, which was detected with $\Delta a$ photometry as well. They noticed that the detection was caused by a slightly lower reddening of the star, but the object belongs to another CP group, the metallic line (Am) stars. Though, a comparable spectroscopic study of the new candidates presented here is more difficult owing to the much fainter magnitudes ($V \sim 15-16$\,mag).  


Figure~\ref{fig:deltaa} shows some more peculiar candidates among the coolest and faintest stars. These correspond to late F-type stars and cannot be considered as classical CP objects. However, in particular the problems in defining the cluster membership makes the detection too uncertain, and we therefore do not list them in Table~\ref{tab:parameters}. Furthermore, in the light of the possible PMS one would expect more stars with negative $\Delta a$ values owing to emission \citep[see discussion by ][]{paunet05}. Spectroscopic data and photometry in the Str{\"o}mgren-Crawford $uvbyH\beta$ system are beneficial to verify the detected candidates, to derive individual reddening values for all cluster stars, and to perform a more detailed membership analysis.

\begin{table*}
\caption{Photometric data of the observed stars. The $\Delta a$ indices are only given for the member stars (m). The complete catalogue is available in electronic form at the CDS.} 
\label{tab:allstars} 
\centering 
\begin{tabular}{l c c c c c c c c c c} 
\hline
ID	& RA  & DEC  & ($g_{1}-y$)  & $\sigma$($g_{1}-y$)  & $a$  & $\sigma(a)$ & $\Delta a$ & ($b-y$) & V & member \\ 
	& [deg] & [deg] & [mag] & [mag] & [mag] & [mag] & [mmag] & [mag] & [mag] & \\ 

\hline 
1 & 202.3031417 & -61.2402489 & 0.298 & 0.010 & 0.102 & 0.009 &  & 0.81 & 14.90 & nm \\
2 & 202.3363500 & -61.2338533 & 0.195 & 0.010 & 0.119 & 0.011 & 68 & 0.64 & 16.90 & m \\
3 & 202.3273333 & -61.2315389 & 0.181 & 0.013 & 0.102 & 0.010 & 54 & 0.62 & 16.74 & m \\
4 & 202.3025208 & -61.2312953 & 0.245 & 0.013 & 0.072 & 0.010 &  & 0.72 & 17.28 & nm \\
5 & 202.2745708 & -61.2310086 & 0.019 & 0.006 & 0.021 & 0.005 & 4 & 0.35 & 14.88 & m \\

\hline 
\end{tabular}
\end{table*}

\section{Summary and conclusions}
\label{summary}
We analysed the very young open cluster Hogg~16, which adds an important contribution to the investigation of the occurrence of CP stars at the earliest stage of their evolution.

We used the DG method by \citet{poehnl10} to estimate the global parameters of 
Hogg~16, however, particularly for the B-type cluster stars we applied the Q-method on available \ubv\ data to estimate the mean reddening $E(B-V)=0.45\pm0.05$\,mag.

The remaining cluster parameters of Hogg~16 are log\,\textit{t} $\leq$ 7.4, 
$(m-M)_0$ = 11.55\,mag, and Z = 0.021 $\pm$ 0.007. 
The parameters agree very well with the results for the neighbouring somewhat more populous cluster Collinder~272, suggesting a physical relationship of the two objects. 
Assuming roughly the same distance for both objects, the angular separation of their centres ($11\arcm$) corresponds to only about 6\,pc. 
The proper motions of the two clusters agree within the errors as well.

The characteristic flux depression of mCP stars at 
$\lambda \sim 5200\,\textup{\AA}$
allows an efficient detection of these objects by the $\Delta a$ photometric system. 
We therefore used this photometric system for an investigation of Hogg~16. The diagnostic $\Delta a$ diagram shows two stars with negative $\Delta a$ indices, which is an indication of emission and the objects can be classified as Be stars. One object (CD\,$-$60\,4701) was also classified as possible Be star using H$\alpha$ photometry. For the other Be star 
candidate (CPD\,$-$60\,4706) there are neither spectral types available in the literature, nor the star shows signs of (photometric) H$\alpha$ emission. More observations, in particular spectroscopic data, are clearly needed owing to the strong spectral variability of Be stars. For another Be star 
in our field of view (CPD\,$-$60\,4703) the available data suggest that the object is most likely a non-member supergiant. 
One B-type star (CD\,$-$60\,4699) is located well above the normality line in the $\Delta a$ diagram and its position in the CMD indicates that it is probably a mCP candidate with 
a mass of about 8\,\msun. Thus, it might belong to the most massive mCP stars, the helium peculiar objects. However, an analysis of spectral data does not confirm this hypothesis.

Three mCP star candidates (No. \,37, 84, and 97) are found in the domain of early F-type stars. One object (No.\,37) is very close to the border of significance and is possibly not a cluster star owing to a lower reddening. However, the other two are probable PMS mCP objects, an assumption based on the position in the CMDs and the found IR-excess. According to their ages,  it is also not clear if they indeed belong to Hogg~16 or are outskirt area members of Collinder~272. Upcoming data releases of the $Gaia$ mission will certainly help to distinguish the two populations more clearly. In any case, the detection of PMS mCP candidates might be a very promising result, because it can impose very strong constraints for the diffusion theory. Though, spectroscopic follow-up observations are needed to confirm their nature.
There are some additional peculiar candidates among the coolest and faintest stars. These would correspond to very late F-type stars and cannot be considered as classical CP objects. 
Additional spectroscopic data are certainly needed to verify also the nature of these objects.

The frequency of mCP stars in very young open clusters appears in general low if inspecting the $\Delta a$ results of other young objects \citep[e.g.][]{paunzen02}. \citet{netopil14b} present a preliminary reanalysis of previous $\Delta a$ studies, which shows a clear dependence of the mCP incidence as a function of the cluster age and a maximum frequency between about 50\,Myr and 200\,Myr. This corresponds to the age range when most mCP stars are on the main-sequence. However, by a comparison with semi-theoretical mCP frequencies, they conclude that mCP stars are already developed as soon as they arrive onto the main sequence or even before.  In each age group, there are cluster to cluster variations of the mCP incidence, suggesting that additional dependencies (e.g. metallicity or galactic location) might be present which favour the formation of the objects. The open cluster NGC~2516, for example, hosts six mCP stars, whereas NGC~6451 is apparently free of mCP objects. Both clusters are very well studied in respect of their mCP star content and are of comparable age ($\sim$\,100 to 200\,Myr). Future data releases of the $Gaia$ satellite mission along with a larger $\Delta a$ cluster sample will certainly provide more details on this topic.

\section*{Acknowledgements}
\small
MN acknowledges the support by the Czech Science Foundation [14-26115P] and we also acknowledge a grant by M\v{S}MT [7AMB17AT030]. OIP and YJA: Visiting Astronomer, Complejo Astron\'omico El Leoncito operated under agreement between the Consejo Nacional de Investigaciones Cient\'ificas y T\'ecnicas de la Rep\'ublica Argentina and the National Universities of La Plata, C\'ordoba and San Juan. 
YJA: This work is within of the subsidy of the National University of La Plata (j\'ovenes investigadores 2016).
This research has made use of several catalogues and tools: The WEBDA database, operated at the Department of 
Theoretical Physics and Astrophysics of the Masaryk University. VOSA, developed under the Spanish Virtual Observatory project supported from the Spanish MICINN through grant AyA2011-24052. Data products from the Wide-field Infrared Survey Explorer, which is a joint project of the University of California, Los Angeles, and the Jet Propulsion Laboratory/California Institute of Technology, funded by the National Aeronautics and Space Administration. Data from the European Space Agency (ESA)
mission {\it Gaia} (\url{http://www.cosmos.esa.int/gaia}), processed by
the {\it Gaia} Data Processing and Analysis Consortium (DPAC,
\url{http://www.cosmos.esa.int/web/gaia/dpac/consortium}). Funding
for the DPAC has been provided by national institutions, in particular
the institutions participating in the {\it Gaia} Multilateral Agreement. We also thank the organizers of the Awareness conference on European Astronomy in the Optical and IR domain: An ESO/Opticon/IAU summer school on modern instruments, their science case, and practical data reduction, which took place in Brno, Czech Republic, in Sept. 2015.

\section*{References}

\bibliography{ho16}

\end{document}